\documentclass[journal=cgdefu,manuscript=article,layout=twocolumn,citeautoscript=false]{achemso}

\usepackage{lipsum}
\usepackage{kantlipsum}
\usepackage{verbatim}
\usepackage{sidecap}
\usepackage[mathlines]{lineno}
\usepackage{amsmath,amssymb,amsfonts}
\usepackage[osf]{newpxtext}
\usepackage{newpxmath}
\usepackage[utf8]{inputenx}
\usepackage{graphicx}
\usepackage[dvipsnames]{xcolor}
\usepackage[pdftex]{hyperref}
\setkeys{acs}{doi = true, articletitle = true}
\usepackage{braket}
\usepackage{xspace}
\usepackage{dblfloatfix}
\usepackage{soul}
\usepackage{chemformula} 
\usepackage[T1]{fontenc} 
\usepackage{siunitx}
\usepackage{float}
\usepackage[normalem]{ulem} 

\DeclareMathAlphabet{\mathcal}{OMS}{cmsy}{m}{n}

\hypersetup{
	pdfauthor={Philipp John},
	bookmarksnumbered=true,
	pdftitle={GaN-on-TiN},
	colorlinks,
	citecolor=black,
	linkcolor=black,
	urlcolor=black}

\author{Philipp\,M.\,John}
\email{john@pdi-berlin.de}
\affiliation
{Paul-Drude-Institut f\"ur Festk\"orperelektronik, Leibniz-Institut im Forschungsverbund Berlin\,e.V., Hausvogteiplatz 5-7, 10117 Berlin, Germany}
\author{Thomas Auzelle}
\altaffiliation{These authors contributed equally to this work.}
\affiliation
{Paul-Drude-Institut f\"ur Festk\"orperelektronik, Leibniz-Institut im Forschungsverbund Berlin\,e.V., Hausvogteiplatz 5-7, 10117 Berlin, Germany}
\author{Pierre-Marie Coulon}
\altaffiliation{These authors contributed equally to this work.}
\affiliation{Universit\'e Côte d’Azur, CRHEA, CNRS, 06905 Sophia-Antipolis Cedex, France}
\author{Blandine Alloing}
\altaffiliation{These authors contributed equally to this work.}
\affiliation{Universit\'e Côte d’Azur, CRHEA, CNRS, 06905 Sophia-Antipolis Cedex, France}
\author{Ileana Florea}
\affiliation{Universit\'e Côte d’Azur, CRHEA, CNRS, 06905 Sophia-Antipolis Cedex, France}
\author{Mikel G\'omez Ruiz}
\author{Jonas L\"ahnemann}
\author{Lutz Geelhaar}
\affiliation
{Paul-Drude-Institut f\"ur Festk\"orperelektronik, Leibniz-Institut im Forschungsverbund Berlin\,e.V., Hausvogteiplatz 5-7, 10117 Berlin, Germany}
\author{Oliver Brandt}
\affiliation
{Paul-Drude-Institut f\"ur Festk\"orperelektronik, Leibniz-Institut im Forschungsverbund Berlin\,e.V., Hausvogteiplatz 5-7, 10117 Berlin, Germany}

\title{Epitaxial integration of GaN \textmu-platelets and thin films on metallic TiN using a hybrid approach}


\begin{document}

	
	
	
	
	
	

	\begin{abstract}
	\textbf{
	We demonstrate the synthesis of epitaxial, self-organized GaN \textmu-platelets on metallic TiN thin films. To this end, we first nucleate self-assembled GaN nanowires with number densities between 0.01 to 300\,\textmu m$^{-2}$ by molecular beam epitaxy. Subsequent overgrowth of these seed nanowires by metal-organic vapor phase epitaxy yields GaN \textmu-platelets with corresponding number densities. The platelets exhibit  N-polarity  and form an ohmic contact to the metallic TiN substrate. By adjusting the growth conditions during the metal-organic vapor phase epitaxy, particularly the carrier gas and precursor ratio, the morphology of the GaN can be controlled across a broad range. Growth under N$_{2}$ carrier gas and with a 1:1 NH$_{3}$/trimethylgallium precursor flow ratio results in platelet ensembles with broad diameter and height distributions. In contrast, employing an N$_{2}$/H$_{2}$ carrier gas mixture promotes the formation of more homogeneous ensembles with narrower size distributions around a diameter of  $\approx$2\,\textmu m. Finally,  N-rich growth conditions under H$_{2}$  carrier gas enhance lateral growth, leading to the formation of coalesced GaN layers. GaN grown under these latter conditions exhibits strong near-band-edge emission, accompanied by only weak deep-level luminescence at room temperature.
}
	\end{abstract}
	\maketitle


\section{Introduction}

Group-III nitrides, particularly GaN, are the dominant semiconductor materials in solid-state lighting  \cite{pust_nat.mater._2015,weinold_nat.energy_2025} and are becoming increasingly important for power electronics. \cite{amano_j.phys.appl.phys._2018,tu_nat.rev.electr.eng._2024}. Currently, common III-nitride devices such as light-emitting diodes (LEDs) and high electron-mobility transistors (HEMTs) are typically based on quasi-vertical and lateral architectures, respectively. A fully vertical architecture, in contrast, uses a conductive substrate as the bottom electrode, thus enabling direct vertical current conduction through the active region of a device. This design offers several advantages over conventional lateral architectures, including superior current spreading, improved heat dissipation, reduced series resistance, and potentially higher integration density. \cite{li_ieeetrans.electrondevices_2012,zhang_j.phys.appl.phys._2018,you_microelectron.reliab._2021,shankari_microelectron.eng._2026} Despite these benefits, vertical III-nitride-based devices remain relatively scarce.

The limited vertical integration of III-nitrides likely stems from two main challenges. First, bulk GaN substrates are still mainly available on a small scale and at high cost. Second, while heteroepitaxy of GaN on large and cost-effective substrates such as sapphire and Si is nowadays widely established, neither substrate offers a straightforward pathway towards vertical GaN-based devices. Due to its electrically insulating nature, sapphire prevents any vertical current transport through the substrate. Growth on Si, on the other hand, generally requires an insulating buffer layer such as AlN to suppress the reaction between Ga and Si known as meltback etching. Consequently, fabricating vertical devices from heterostructures grown on Si demands additional and complex processing steps, such as layer transfer or etching through both the Si substrate and the insulating buffer layer to establish an electrical backside contact \cite{zhang_j.phys.appl.phys._2018}.

\begin{figure}[t]
	\includegraphics[width=1.0\columnwidth]{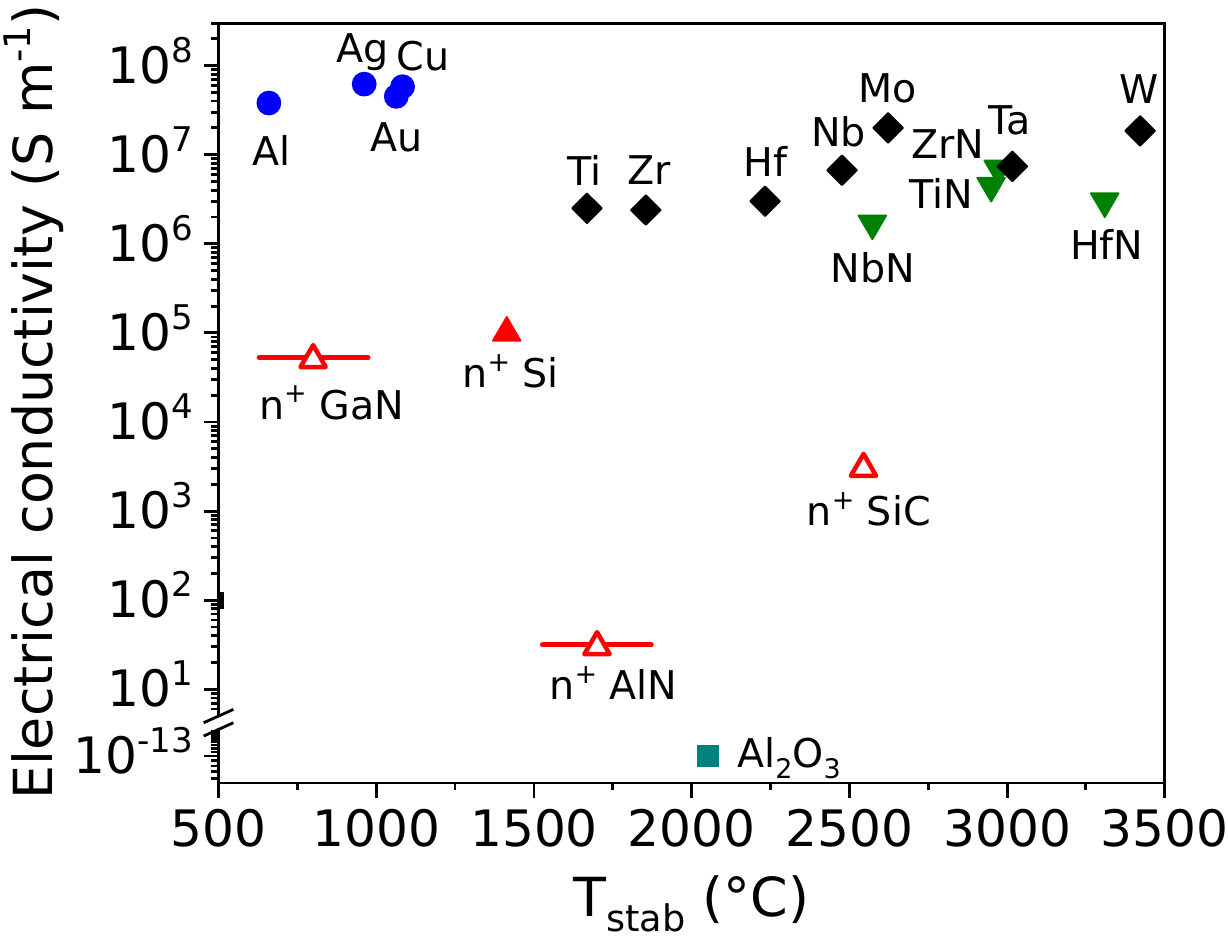}
	\caption{Electrical conductivity of possible substrate materials for III-nitride growth, plotted against their thermal stability limit T$_{\mathrm{stab}}$, i.\,e., their melting point (filled symbols) or sublimation temperature (open symbols). Materials include bulk semiconductors (\textcolor{red}{$\blacktriangle$}) and insulators (\textcolor{cyan!60!black}{$\blacksquare$}), as well as thin films of noble metals (\textcolor{blue}{\scalebox{1.2}{$\bullet$}}), transition metals (\textcolor{black}{\scalebox{1.5}[1]{$\blacklozenge$}}) and transition metal nitrides (\textcolor{green!60!black}{$\blacktriangledown$}). Electrical conductivity data are taken from Refs.\,\citenum{chapman_j.appl.phys._1963,champion_br.j.appl.phys._1964,mitchel_j.electron.mater._1997,adachi_j.alloyscompd._2005,konczewicz_semicond.sci.technol._2022a,gablech_adv.electron.mater._2023a,holleman__2016}, values for T$_{\mathrm{stab}}$ are taken from Refs.\,\citenum{schneider_j.res.natl.bur.stand.sect.phys.chem._1967,groh_phys.statussolidia_1974,olesinski_bull.alloyphasediagr._1984,pisch_j.cryst.growth_1998,holleman__2016,peters_j.appl.phys._2023}.}
	\label{fig0}
\end{figure}

An interesting alternative to classical semiconducting substrates are metallic thin films, exhibiting orders of magnitude higher electrical conductivity, as displayed in Figure\,\ref{fig0}. Furthermore, they can be deposited on low-cost substrates of any size, thus presenting a cost-efficient, scalable approach. In optoelectronic applications, a metallic substrate can additionally serve as a back reflector, potentially enhancing light extraction by reflecting photons emitted towards the metal layer back to the top surface. 

Yet, classical conductors such as Ag, Cu, Al and Au exhibit relatively low melting points, not suitable for the growth of epitaxial III-nitrides taking place at elevated temperatures (cf. Figure\,\ref{fig0}). In contrast, transition metals and, especially, their nitrides exhibit an outstanding thermal stability and, at the same time, a high electrical conductivity. 

Various refractory metals and transition metal nitrides  have been explored as substrates for GaN growth, predominantly by molecular beam epitaxy (MBE). Examples include Ti, Ta and Mo thin films  \cite{wolz_nanolett._2015,sarwar_small_2015} and foils \cite{zhao_nanolett._2016,may_appl.phys.lett._2016,calabrese_appl.phys.lett._2016}, as well as, more recently, ZrN thin films \cite{olszewski_nanomaterials_2023,tiagulskyi_nanoscale_2025}. However, these studies have focused on the growth of self-assembled GaN nanowires, generally leading to nanowire ensembles with high number densities and, when grown on polycrystalline substrates, often broad distributions of in-plane orientations. Although vertical electrical conduction and even nanowire-based LEDs have been demonstrated using such ensembles \cite{sarwar_small_2015,zhao_nanolett._2016,may_appl.phys.lett._2016,tiagulskyi_nanoscale_2025}, the limited control over nanowire density, morphology and orientation remains a challenge for their integration into other devices.

In our previous works, we have demonstrated the growth of III-nitride nanowires on epitaxial TiN yielding 
self-assembled GaN \cite{wolz_nanolett._2015,auzelle_nanotechnology_2023} and AlN  \cite{azadmand_phys.statussolidirrl-rapidres.lett._2020,john_nanotechnology_2023} nanowire ensembles with well-defined in-plane orientations. 
Furthermore, we recently demonstrated precise control over the GaN nucleation process by depositing SiN$_{x}$ islands that serve as preferential nucleation sites, allowing the nanowire density to be tuned over several orders of magnitude \cite{auzelle_nanotechnology_2023}. Although the integration of bottom-up nanostructures into electronic devices is generally challenging, our TiN-supported III-nitride nanowire ensembles have recently been incorporated into vertical piezoelectric nanogenerators \cite{buatip_acsappl.nanomater._2024,notarangelo__2026}. This demonstration highlights the potential of TiN as an integrated bottom electrode for other III-nitride-based devices.

In contrast to the self-assembled nanowire  growth typically obtained by MBE, metal-organic vapor-phase epitaxy (MOVPE) of GaN on TiN has been reported to result in the nucleation of large, $\mu$m-sized 3D crystallites, as found in several studies by the group of Murakami \cite{watanabe_mater.trans._2005,ito_mrsproc._2006,uchida_j.electron.mater._2006,ito_j.electron.mater._2009}. The formation of such islands has been attributed to the poor wetting of GaN on TiN, potentially associated with a high GaN/TiN interfacial energy. Although these 3D crystallites can subsequently be coalesced into GaN layers by lateral overgrowth, the resulting layers exhibit root-mean-square (RMS) surface roughnesses of several hundred nanometers. The deposition of an AlN buffer layer has been recently demonstrated to yield smoother GaN layers on TiN \cite{horich_front.mater._2025}. However, the introduction of such an insulating buffer layer again compromises vertical electrical conduction.

In this work, we combine the advantages of MBE and MOVPE in a hybrid approach that decouples GaN nucleation from subsequent lateral growth. We first employ MBE to nucleate self-assembled nanowires on 200\,nm-thick TiN layers, similar to our previous work \cite{auzelle_nanotechnology_2023}. These seed nanowires are subsequently overgrown by MOVPE to form both $\mu$m-sized GaN platelets and coalesced layers, thereby shifting the growth front from the GaN/TiN interface towards the established GaN surfaces. Such an MBE/MOVPE hybrid approach has been recently introduced by Henriques et al. \cite{henriques_cryst.growthdes._2024,henriques_cryst.growthdes._2025}, but on insulating 300\,nm SiO$_{2}$/Si substrates and with patterned graphene patches serving as nucleation sites.
Our GaN/TiN heterostructures exhibit an ohmic contact to the metallic TiN layer even after the high-temperature overgrowth, demonstrating that the electrical functionality of the TiN substrate is retained. The $\mu$-platelets are thus promising building blocks for future  vertical  $\mu$-LED architectures, while coalesced GaN layers could enable novel vertical GaN power devices.

\section{Experimental details}
\label{sec:experiments}

GaN nanowires are grown by MBE at $\approx$840\,$^\circ{}$C on 200\,nm-thick sputtered TiN(111)/Al$_{2}$O$_{3}$(0001) templates, as described in our previous works \cite{auzelle_nanotechnology_2023,john_nanotechnology_2023,john_nanolett._2024}. The nanowire nucleation does not occurr directly on the TiN thin film, but exclusively on SiN$_{x}$ islands deposited in the MBE chamber prior to GaN growth. Consequently, the nanowire number density is directly proportional to the density of the SiN$_{x}$ islands, which can be precisely controlled through the amount of deposited SiN$_{x}$ \cite{auzelle_nanotechnology_2023}. 

Self-assembled nanowire ensembles of different densities (0.01\,\textmu m$^{-2}$, 0.2\,\textmu m$^{-2}$, 5\,\textmu m$^{-2}$ and 300\,\textmu m$^{-2}$) are simultaneously overgrown by MOVPE for 10\,min, using trimethylgallium (TMGa) and NH$_{3}$ as precursors. A total carrier gas flow of 8\,slm of either N$_{2}$, H$_{2}$ or an N$_{2}$/H$_{2}$ mixture is used. The growth temperature is kept between 925 and 1020\,$^\circ{}$C and the pressure between 75 and 100\,Torr. In some growth experiments, 0.01\,sccm of SiH$_{4}$ is added, in order to achieve \textit{n}-type doping.


The morphology of the grown GaN structures is analyzed with scanning electron microscopy (SEM) and atomic force microscopy (AFM). Their polarity is assessed by wet etching experiments, using a 20\,\% KOH solution for 3\,min at 70\,$^\circ{}$C. 

To investigate the microstructure of selected samples, scanning transmission electron microscopy (STEM) was performed in a Titan SPECTRA 200 (ThermoFisher), equipped with a cold field emission gun and a Cs aberration probe corrector. Cross-section lamellae were prepared by focused ion beam sectioning inside a Thermo Fisher Helios 5 UX SEM. 

The electrical contact between the grown GaN and the TiN layer was investigated by current–voltage measurements through several GaN platelets and the TiN substrate using a conventional two-point measurement probe station, with one probe needle placed on a $\mathrm{\mu m}$-sized GaN platelet and the other on the TiN layer.  

Optical characterization was performed by cathodoluminescence (CL) spectroscopy at room temperature, using a Delmic Sparc Spectral CL system particularly designed for the detection of UV emission. Excitation was performed inside a Thermo Fisher Verios 5 UC field emission SEM, using an acceleration voltage and beam current of 3\,kV and 0.4\,nA, respectively. The emitted light was collected by a parabolic mirror and focused on the 10\,$\mathrm{\mu m}$ wide entrance slit of the spectrometer. Signal recording takes place by a UV-enhanced CCD camera at the output of the spectrometer. At least five different platelets were analyzed and representative spectra are shown.

\section{Results and Discussion}

\begin{figure}
	\includegraphics[width=1.0\columnwidth]{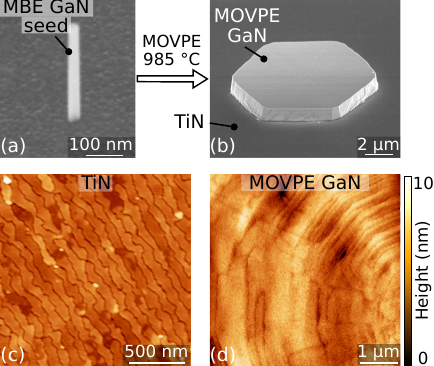}
	\caption{MBE-grown GaN nanowire on sputtered TiN thin film (a) before and (b) after lateral overgrowth by MOVPE, yielding an $\mathrm{\mu m}$-sized hexagonal platelet. (c) and (d) show atomic force topographs of the TiN surface and the GaN $\mathrm{\mu}$-platelet after the MOVPE growth, respectively. The height scale is identical for both topographs and shown in (d).}
	\label{fig1}
\end{figure}

Figure\,\ref{fig1} displays a typical overgrowth of MBE-grown GaN nanowires by MOVPE. In particular, a nanowire ensemble with a number density of 0.01\,\textmu m$^{-2}$ [Figure\,\ref{fig1}(a)] self-assembled on TiN is overgrown at 985\,\,$^\circ{}$C under pure N$_{2}$ carrier gas, using 40\,sccm TMGa and 40\,sccm NH$_{3}$. The MOVPE overgrowth leads to both a lateral and axial expansion of the seed nanowire, forming a $\mathrm{\mu m}$-sized hexagonal GaN platelet [Figure\,\ref{fig1}(b)]. Notably, the TiN(111) surface remains intact after MOVPE growth, exhibiting flat grains arranged in long, parallel domains with an average diameter of $\approx$75\,nm and a root mean square (RMS) roughness of 1.4\,nm, as revealed by the atomic force topograph in Figure\,\ref{fig1}(c). This morphology contrasts the initial TiN surface that is characterized by small circular grains with inclined \{100\} facets, and a grain diameter of $\approx$25\,nm, as investigated in our previous works \cite{auzelle_nanotechnology_2023,john_nanotechnology_2023}. The high temperature during the MOVPE thus leads to a structural reshaping of the TiN with significant grain growth and the formation of flat \{111\} facets. Note that this process may already commence during the initial heating phase and may be further promoted by the presence of NH$_3$, which has been reported to enhance recrystallization in TiN \cite{sun_appl.phys.lett._1996}. The parallel domain orientation found here may be kinetically governed by the atomic step morphology of the Al$_{2}$O$_{3}$(0001) substrate. Note that a similar morphology has been observed by \citet{wright_phys.rev.mater._2021} after the MBE growth of $\delta$-NbN(111) on SiC(0001), where alternating domains have been identified as the two in-plane rotational domains of the NbN. Such domains inevitably form during the epitaxy of cubic rock-salt materials with (111) orientation on hexagonal wurtzite (0001) surfaces due to the symmetry mismatch between the two surfaces, exhibiting $3m$ and $6mm$ symmetry, respectively \cite{grundmann_phys.statussolidib_2011}.

\begin{figure*}[t]
	\includegraphics[width=1\textwidth]{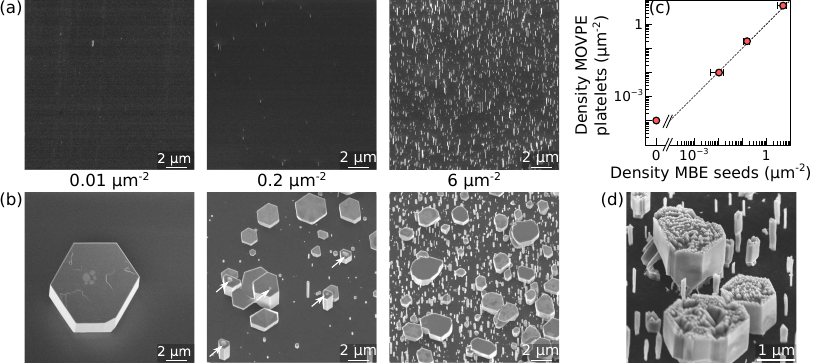}
	\caption{GaN nanowire seeds with number densities  of 0.01 (left) 0.2 (center) and 5\,\textmu m$^{-2}$ (right) before (a) and after (b) MOVPE overgrowth at 985\,$^\circ{}$C under N$_{2}$ carrier gas and the presence of SiH$_{4}$. (c) Number density of MOVPE-grown GaN $\mathrm{\mu}$-platelets as a function of MBE seed nanowire density. Note that for the highest seed density (5\,\textmu m$^{-2}$), both large platelets and residual thin nanowires are included in the count. (d) GaN $\mathrm{\mu}$-platelets with a number density of 5\,\textmu m$^{-2}$ after KOH etching. White arrows in (b) indicate Ga-polar inversion domains.}
	\label{fig2}
\end{figure*}


Figure\,\ref{fig1}(d) illustrates the morphology of a selected area of an hexagonal GaN $\mathrm{\mu}$-platelet. The atomic force topograph of this platelet reveals a relatively smooth surface (RMS = 1.0\,nm) with a growth mode characterized by step-bunching. Note that the surface morphology of the GaN $\mathrm{\mu}$-platelets is generally found to be dependent on the MOVPE growth conditions and can vary even within one sample, as further revealed below.


	\begin{figure*}[t] 
	\centering
	\begin{minipage}[c]{0.75\textwidth}
		\includegraphics[width=\textwidth]{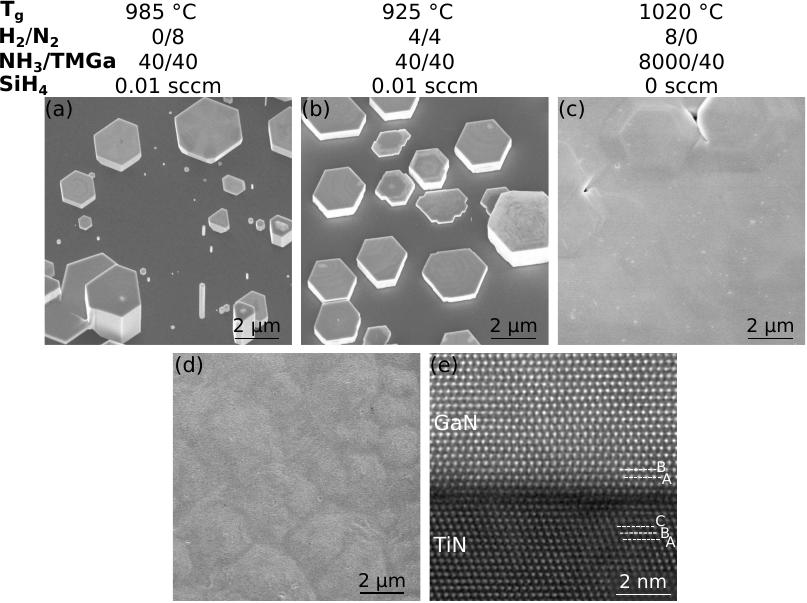}
	\end{minipage}%
	\hfill
	\begin{minipage}[c]{0.23\textwidth}
		\caption{
			(a)-(c) GaN nanowires with a number density of 0.2\,\textmu m$^{-2}$ and overgrown by MOVPE with different growth conditions, as indicated above. (d) MOVPE overgrowth of GaN nanowires with a number density of 300\,\textmu m$^{-2}$, with the growth conditions used in (c). (e) High-resolution HAADF micrograph of the GaN/TiN interface from the sample shown in (d).
		}
		\label{fig3}
	\end{minipage}
\end{figure*}


Next, we investigate the impact of the GaN seed nanowire number density on the resulting platelet morphology. Figure\,\ref{fig2}(a) displays secondary electron micrographs of three GaN nanowire ensembles with densities of 0.01 (left panel), 0.2 (center panel) and 5\,\textmu m$^{-2}$ (right panel) grown by MBE. These samples are overgrown by MOVPE in the same growth run, using similar conditions to the ones used in Figure\,\ref{fig1}. In addition, SiH$_{4}$ is introduced during growth to achieve \textit{n}-type doping, thereby enabling the electrical measurements presented later.

The result of the MOVPE overgrowth is displayed in Figure\,\ref{fig2}(b). The secondary electron micrographs evidence the formation of GaN platelets for all overgrown nanowire ensembles, yet, with distinct morphologies. The overgrowth of the low-density nanowire ensemble (0.01\,\textmu m$^{-2}$) yields few and very large GaN platelets (left panel). An increase in nanowire number density to 0.2\,\textmu m$^{-2}$ leads also to an increased platelet density, but with a smaller average diameter and height (center panel) and broad diameter and height distributions. This trend is continued for a seed nanowire density of 5\,\textmu m$^{-2}$ (right panel). The quantitative diameter and height distributions of all GaN $\mathrm{\mu}$-platelet ensembles are displayed in Figure\,S1 of the supporting information. 

Figure\,\ref{fig2}(c) displays the number density of MOVPE platelets resulting from the overgrowth, as a function of initial MBE-grown GaN seed nanowire density. Note that for the highest seed density (5\,\textmu m$^{-2}$, cf.  Figure\,\ref{fig2}(b), right panel)  the reported platelet density includes both large $\mathrm{\mu}$-platelets and residual thinner nanowires that did not form full platelets during overgrowth. 
The dependence is close to a 1:1 ratio (dotted line), with negligible parasitic growth found on the bare TiN substrate in absence of any seed nanowires.  This result confirms that the TiN layer acts as a mask for GaN nucleation during the MOVPE process, the growth thus only taking place in the presence of MBE seed nanowires, leading to their axial and radial enlargement. 

The decrease of platelet diameter and height as a function of seed nanowire density may be explained by the precursor diffusion processes characterizing selective-area MOVPE, as described in previous studies \cite{olsson_j.cryst.growth_2006,dubrovskii_nanomaterials_2022,stepniak_mater.sci.eng.b_2026}. In particular, when the distance between initial seed nanowires becomes smaller than the effective diffusion length of the precursor molecules, a decrease of the local precursor concentration adjacent to the growing structures is expected, thus lowering the overall platelet growth rate. Consequently, the local precursor concentration close to a nanowire or platelet becomes highly dependent on the neighboring mask area, possibly explaining the large variation of growth rate also within each self-assembled nanowire ensemble, since each nanowire exhibits a characteristic nearest-neighbor distance. 


In order to investigate the polarity of our GaN $\mathrm{\mu}$-platelets, we performed KOH etching experiments. As displayed in Figure\,\ref{fig2}(d), the KOH solution leads to an etching of the top facets in the MOVPE ensemble with an initial GaN seed nanowire density of 5\,\textmu m$^{-2}$ (cf. Figure\,\ref{fig2}(b), right panel). This result evidences the N-polarity of our $\mathrm{\mu}$-platelets \cite{zuniga-perez_appl.phys.rev._2016}, consistent with the polarity of the GaN seed nanowires assessed independently by piezoelectric force microscopy \cite{jaloustre_acsappl.nanomater._2021}. The occasional formation of small, 3D islands on the (000\={1}) top facets (indicated by the white arrows in Figure\,\ref{fig2}(b), center panel) likely indicates the formation of Ga-polar inversion domains within the N-polar matrix, as found previously by \citet{henriques_cryst.growthdes._2025} for their MBE-MOVPE grown GaN $\mathrm{\mu}$-platelets. Consistently, these areas are not etched by the KOH solution (not shown).

To improve the homogeneity of our self-assembled GaN $\mathrm{\mu}$-platelets, we next investigate the impact of MOVPE growth conditions on the overgrowth of the GaN nanowire ensemble with the number density of 0.2\,\textmu m$^{-2}$. 
Figure\,\ref{fig3}(a) displays again the overgrowth ensemble  obtained in Figure\,\ref{fig2}, center panel, i.\,e., grown under pure N$_{2}$ carrier gas  at 985\,$^\circ{}$C and 0.01\,sccm SiH$_{4}$, exhibiting a large diameter and height distribution. 

In contrast, Figure\,\ref{fig3}(b) displays an overgrowth of the same GaN seed nanowires, but with a 1:1 mixture of N$_{2}$ and H$_{2}$ carrier gases, carried out at a substrate temperature of 925\,$^\circ{}$C and in presence of 0.01\,sccm SiH$_{4}$. The resulting GaN $\mathrm{\mu}$-platelets have a larger average diameter compared to those grown under pure N$_{2}$ carrier gas [cf. Figure\,\ref{fig3}(a)], indicating that these conditions favor besides axial also lateral overgrowth. The platelet's diameter distribution is with $(1.9 \pm 0.4)$\,\textmu m the narrowest among all grown ensembles. 

Finally, Figure\,\ref{fig3}(c) displays an MOVPE overgrowth carried out under pure H$_{2}$ carrier gas and conditions typical for GaN layer growth, i.e., a growth temperature of 1020\,$^\circ{}$C, a TMGa flow of 60\,sccm and an NH$_{3}$ flow of 8\,slm, under a pure H$_{2}$ carrier gas atmosphere and in absence of SiH$_{4}$. As a result, strong lateral growth is found, leading to the formation of an almost entirely coalesced GaN layer. The RMS roughness of the GaN thin film is with $\approx$110\,nm very high, likely due to a large seed nanowire length distribution, which is related to subsequent nucleation of nanowires during the long incubation times occurring in our GaN nanowires \cite{auzelle_nanotechnology_2023}. 

To further enhance coalescence and decrease surface roughness, we attempted an overgrowth on short, highly dense nanowires (ensemble with an average length of ($47 \pm 15$)\,nm and a number density of 300\,\textmu m$^{-2}$), the morphology being displayed in Figure\,S2 of the supporting information.  Although this approach resulted in the formation of a fully coalesced thin film, [see Figure\,\ref{fig3}(d)] the obtained GaN layer still exhibits a high RMS roughness of approximately $\approx$50\,nm. Hence, growth on nanowire ensembles with narrower length distributions and/or an increase of overgrowth thickness is necessary to further reduce the surface roughness of such coalesced GaN layers on TiN.

We performed STEM measurements to investigate the interface between the overgrown GaN and the TiN template. A sharp interface with no evident inter-diffusion is revealed in the high-resolution high-angle annular dark field (HAADF) micrograph in Figure\,\ref{fig3}(e). The GaN layer exhibits an \textit{AB} stacking sequence  along the [11\={2}0]$_{\mathrm{GaN}}$ zone axis, characteristic for the hexagonal wurtzite structure. In contrast, the TiN layer exhibits an \textit{ABC} stacking sequence along the [1\={1}0]$_{\mathrm{TiN}}$ zone axis, consistent with its cubic rock-salt structure.   The complete epitaxial relationship is thus GaN(000\={1})[11\={2}0]$||$ TiN(111)[1\={1}0], as commonly observed for group-III nitrides grown on TiN \cite{uchida_j.electron.mater._2006,wolz_nanolett._2015}.
Note that the interface of the heterostructure locally exhibits an increased roughness (not shown), which can be attributed to the surface roughness of the TiN layer [cf. Figure\,\ref{fig1}(c)].

X-ray diffraction measurements reveal relatively narrow rocking curves. A full-width at half maximum (FWHM) of 0.57$^\circ{}$ is found for an $\omega$-scan across the (000\={2}) reflection, much narrower than the corresponding FWHM of the initial seed nanowire sample of 2.1$^\circ{}$, as typically found due to a strong nanowire-to-nanowire tilt (see Figure\,S2 of the supporting information). This phenomenon has been observed previously during the MOVPE coalescence of GaN nanostructures and is attributed to a reorientation of the initially misoriented, individual GaN nanowires upon the early stage of their coalescence, thus forming a coalesced layer with comparatively higher crystalline quality \cite{wehbe_j.appl.phys._2024}.


\begin{figure}[t]
	\includegraphics[width=0.95\columnwidth]{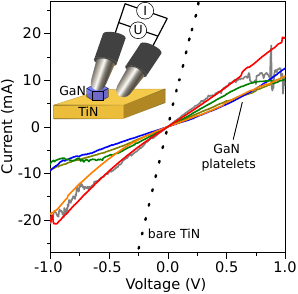}
	\caption{Current-voltage curves measured through several Si-doped GaN platelets and the TiN substrate. The measurements were performed using probe needles, with one needle placed on a large GaN platelet and the other one on the TiN layer, as indicated in the top left corner. The dashed line corresponds to a reference measurement with both needles placed on the TiN surface. Different solid lines correspond to measurements on different GaN platelets.}
	\label{fig4}
\end{figure}


We have demonstrated that the MOVPE growth conditions, particularly the carrier gas mixture and precursor ratio, can be utilized to form either GaN $\mathrm{\mu}$-platelets or coalesced thin films on metallic TiN templates when overgrowing GaN seed nanowires. Yet, for a potential use of our epitaxial GaN-on-TiN platform in functional devices, the evaluation of the electrical contact characteristics between the TiN layer and GaN is essential. For this purpose, we measured two-point probe current-voltage curves. First, we evaluated the conductivity of the TiN substrate after the MOVPE overgrowth by placing both probe needles on the exposed TiN substrate between large GaN platelets. The dotted line in Figure\,\ref{fig4} shows a strong, linear increase in current as a function of voltage, confirming the metallic behavior of the TiN layer after the MOVPE overgrowth. Next, we repeated the same measurement, but with one probe needle placed on a large, Si-doped GaN $\mathrm{\mu}$-platelet (cf. Figure\,\ref{fig2}(b), left panel) and the second one remaining on the TiN layer. These measurements were performed without depositing electrical contacts on the GaN platelets, as schematically displayed in the inset of Figure\,\ref{fig4}. The solid lines display the resulting current-voltage curves through several single GaN $\mathrm{\mu}$-platelets. A quasi-linear relationship is found with a maximum current of $\approx$ 20\,mA at 1.0\,V, indicating an ohmic contact between the GaN and the TiN layer. This ohmic behavior is consistent with previous experimental studies \cite{jeon_appl.phys.lett._2009,zhu_appl.surf.sci._2019} and can be explained by the low work function of TiN \cite{zhu_j.vac.sci.technol.vac.surf.films_2018}.
Note that the measured current, limited by the platelet's lateral dimension (diameter $\leq$10$\mathrm{\mu}$m), is sufficient to power GaN-based $\mathrm{\mu}$-LEDs \cite{li_opt.express_2019,zhang_j.appl.phys._2025} or vertical GaN diodes \cite{zhang_appl.phys.lett._2017}, as such devices typically have comparable or larger contact areas.


\begin{figure*}[t] 
	\centering
	\begin{minipage}[c]{0.75\textwidth}
		\includegraphics[width=\textwidth]{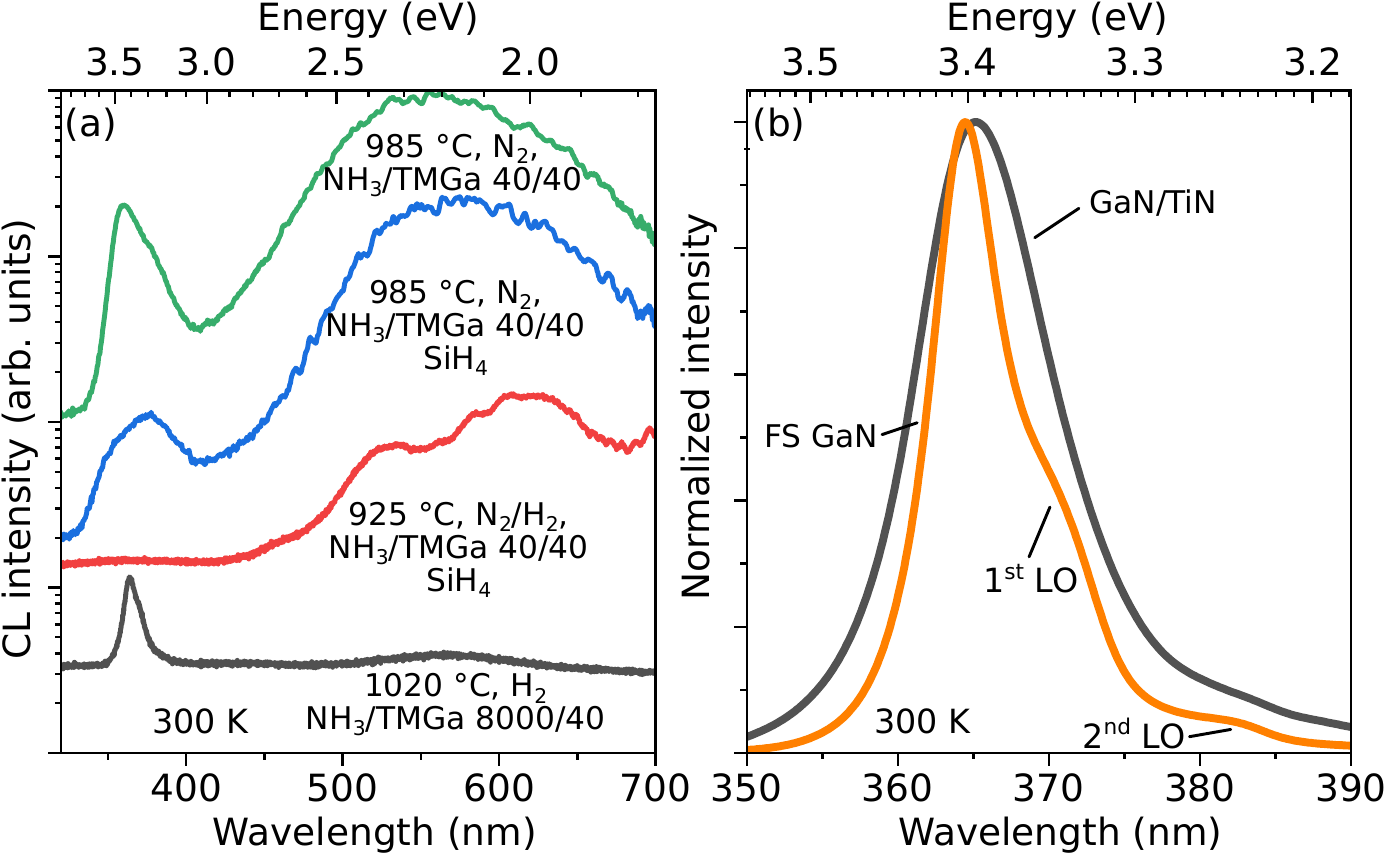}
	\end{minipage}%
	\hfill
	\begin{minipage}[c]{0.24\textwidth}
		\caption{
			(a) CL spectra of GaN nanowires with a number density of 0.2\,\textmu m$^{-2}$, overgrown by MOVPE under different conditions, as indicated below each spectrum. (b) Normalized near-band edge region of the bottom spectrum shown in (a), together with a photoluminescence spectrum of free-standing (FS) GaN for reference. 
		}
		\label{fig5}
	\end{minipage}
\end{figure*}


Finally, we evaluated the optical properties of our GaN platelets and thin films using CL spectroscopy. Figure\,\ref{fig5}(a) displays room-temperature CL spectra of the nanowire ensembles overgrown at different conditions, with an initial number density of 0.2\,\textmu m$^{-2}$. The quasi-coalesced GaN layer grown under H$_{2}$ carrier gas at 1020\,$^\circ{}$C [cf.\,Figure\,\ref{fig3}(c)] is characterized by a dominant near-band edge emission centered around 3.395\,eV, and comparatively weak  emission from deep levels between 2.0 and 2.5\,eV (gray curve, bottom spectrum), indicating a relatively low density of point defects. The near-band edge region of this sample is shown in Figure\,\ref{fig5}(b), together with a reference photoluminescence (PL) spectrum of a free-standing GaN crystal (orange curve), featuring a narrow free exciton transition (FX$_{A}$) at 3.401\,eV (364.4\,nm) and well-defined longitudinal optical (LO) phonon replica. Compared to the narrow peak of the free-standing GaN with an FWHM of 67\,meV, the emission band of our GaN/TiN is broader (FWHM of 100\,meV). This broadening is likely caused by a higher excitation density in CL compared to PL \cite{damilano_j.appl.phys._2024}, the presence inhomogeneous strain in the GaN/TiN layer \cite{kaganer_j.phys.appl.phys._2015}, and, potentially, a higher background doping density \cite{schubert_appl.phys.lett._1997}.

Changing the growth conditions to favor the formation of GaN $\mathrm{\mu}$-platelets drastically alters the emission characteristics. Remarkably, the $\mathrm{\mu}$-platelets with the narrowest height and diameter distribution [cf. Figure\,\ref{fig3}(b)], grown under an H$_{2}$/N$_{2}$ mixture with a 1:1 NH$_{3}$/TMGa precursor flow ratio at 925\,$^\circ{}$C exhibit no near-band luminescence (Figure\,\ref{fig5}(a), red curve). Instead, a pronounced defect band with two broad peaks centered around 2.0\,eV (620\,nm) and 2.35\,eV (525\,nm) is found. In undoped GaN, deep-level luminescence is commonly associated with the presence of C impurities substituting N (C$_{N}$), nitrogen vacancies (V$_{N}$) or oxygen-containing defect complexes \cite{reshchikov_phys.statussolidib_2023}. The formation of C$_{N}$ and V$_{N}$ defects is generally favored under Ga-rich conditions and  at relatively low growth temperatures \cite{ciarkowski_materials_2019,li_rscadv._2020}, in agreement with the conditions used for the overgrowth, and may therefore also contribute to the observed emission. In addition, the presence of SiH${4}$ during the growth of this $\mathrm{\mu}$-platelet ensemble suggests that Si-related defect complexes likely contribute to the near-band edge quenching and to the observed deep-level luminescence \cite{schubert_appl.phys.lett._1997,kaufmann_phys.rev.b_1999,diaz-guerra_j.appl.phys._2003}. 


Similar deep-level luminescence features, this time centered around 2.2\,eV (560\,nm), are found for the GaN $\mathrm{\mu}$-platelets exhibiting large diameter and height distributions [grown at 985\,$^\circ{}$C with a 1:1 NH$_{3}$/TMGa precursor flow ratio under N$_{2}$ carrier gas, cf. Figure\,\ref{fig3}(a)], as displayed in the third-lowest spectrum (blue curve). In addition, a weak, broad near-band edge emission feature is found centered at 3.35\,eV, which may be explained by the slightly higher growth temperature used for this ensemble. The peak position and large FWHM still suggest a very high point defect density.

To investigate the impact of Si doping on the luminescence properties of our GaN $\mathrm{\mu}$-platelets, we grew one ensemble under identical conditions as that of Figure\,\ref{fig3}(a) [blue curve, third-lowest spectrum in Figure\,\ref{fig5}(a)], but without the addition of SiH$_{4}$. The CL spectrum of this undoped ensemble [green curve, top spectrum in Figure\,\ref{fig5}(a)] exhibits spectral features similar the Si-doped one, i.\,e. a pronounced defect band around 2.2\,eV and a broad near-band edge emission band around 3.35\,eV. Yet, the near-band edge luminescence in the undoped sample shows a narrower FWHM, indicating that Si incorporation indeed contributes to the broadening of the near-band edge luminescence. 
\section{Conclusion}
We have demonstrated the integration of epitaxial GaN on metallic TiN thin films through a two-step growth approach, involving the initial nucleation of self-assembled GaN nanowires by MBE, followed by their overgrowth by MOVPE. By tuning the seed nanowire density and the MOVPE overgrowth conditions, the morphology of the resulting GaN/TiN heterostructures can be controlled from relatively homogeneous, micrometer-sized platelets to fully coalesced thin films. The ohmic contact formed between GaN and TiN makes these heterostructures promising templates for the development of vertical III-nitride devices. In particular, the \textit{n}-type GaN platelets, with diameters of approximately 2\,$\mathrm{\mu}$m, offer a potential platform for the fabrication of vertical In$_{x}$Ga$_{1-x}$N based $\mathrm{\mu}$-LEDs. On the other hand, fully coalesced GaN layers on metallic TiN may provide a suitable platform for future vertical power devices and bulk acoustic wave resonators.

	\begin{acknowledgement}
The authors thank their PDI colleagues C. Matzeck for TEM sample preparation and C. Stemmler for dedicated MBE maintenance. They further acknowledge A. Courville from CRHEA for assistance in MOVPE growth. We are grateful to J. Kang for critically reading the manuscript. This work was funded by Deutsche Forschungsgemeinschaft (568879259) and Agence Nationale de la Recherche (ANR-25-CE24-0472) through the project MINT and supported by COST Action OPERA from the European Network for Innovative and Advanced Epitaxy, CA20116 (www.cost.eu). We further acknowledge co-funding by the European Regional Development Fund (ERDF) through the Berlin Senate for the application laboratory time-resolved cathodoluminescence spectroscopy (ZALKAL).
	\end{acknowledgement}
	


	\bibliography{Paper_VPE_regrowth}

\clearpage
\appendix

\setcounter{section}{0}
\setcounter{figure}{0}
\setcounter{table}{0}
\setcounter{equation}{0}
\setcounter{page}{1} 

\renewcommand{\thesection}{S\arabic{section}}
\renewcommand{\thefigure}{S\arabic{figure}}
\renewcommand{\thetable}{S\arabic{table}}
\renewcommand{\theequation}{S\arabic{equation}}

\onecolumn
\section{Supporting information}

\begin{figure*}
	\centering
	\includegraphics[width=0.4\textwidth]{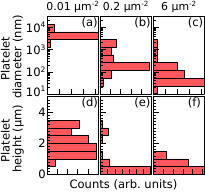}
	\caption{Diameter (a)–(c) and height (d)–(f) statistics of MOVPE-grown GaN $\mathrm{\mu}$-platelets. Each pair of vertically stacked panels corresponds to the same GaN nanowire seed number density, as indicated above.
	}
	\label{figS1}
\end{figure*}

\begin{figure*}
	\includegraphics[width=1\textwidth]{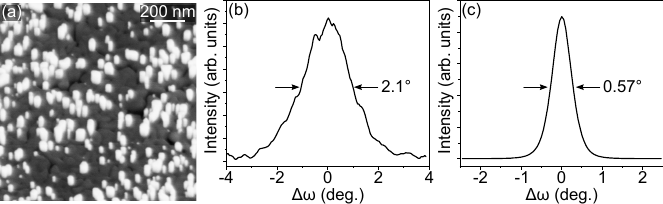}
	\caption{XRD $\omega$-scans of (a) MBE-grown, self-assembled GaN nanowire seeds with a number density of 300\,\textmu m$^{-2}$ and (b) the same nanowire ensemble after MOVPE overgrowth, yielding a coalesced GaN layer.  The MOVPE was performed under H$_{2}$ carrier gas using 40\,sccm TMGa and 8000\,sccm NH$_{3}$ at 1020\,$^\circ{}$C. The rocking curves were measured with a Panalytical X'Pert diffractometer equipped with a two-bounce Ge(220) hybrid monochromator, providing CuK$\alpha _{1}$ radiation. The measurements were performed in double-crystal geometry, with a 1\,mm receiving slit placed in front of the x-ray detector.}
	
	\label{figS2}
\end{figure*}

\end{document}